\documentclass[12pt]{article}
\usepackage{amsfonts,bm}
\usepackage{graphicx}
\baselineskip
\offinterlineskip
\begin{document}

\begin{center}
{\bf Exponential of a Matrix, a Nonlinear Problem and Quantum Gates}
\end{center}

\begin{center}
{\bf  Willi-Hans Steeb$^\dag$ and Yorick Hardy$^\ast$} \\[2ex]

$\dag$
International School for Scientific Computing, \\
University of Johannesburg, Auckland Park 2006, South Africa, \\
e-mail: {\tt steebwilli@gmail.com}\\[2ex]

$\ast$
Department of Mathematical Sciences, \\
University of South Africa, Johannesburg, South Africa, \\
e-mail: {\tt hardyy@unisa.ac.za}\\[2ex]
\end{center}

\strut\hfill

{\bf Abstract.} We describe solutions of the matrix equation
$\exp(z(A-I_n))=A$, where $z \in {\mathbb C}$.
Applications in quantum computing are given.
Both normal and nonnormal matrices are studied.
For normal matrices, the Lambert W-function plays a central role.

\strut\hfill

\section{Introduction}

The motivation of this paper comes from the following observation. 
Let $\sigma_1$, $\sigma_2$, $\sigma_3$ be the Pauli spin matrices
$$
\sigma_1 = \pmatrix { 0 & 1 \cr 1 & 0 }, \quad
\sigma_2 = \pmatrix { 0 & -i \cr i & 0 }, \quad
\sigma_3 = \pmatrix { 1 & 0 \cr 0 & -1 }
$$
and $\sigma_0 \equiv I_2$ be the $2 \times 2$ identity matrix, then
$$
\exp\left(-\frac12 i\pi(\sigma_j-I_2)\right) \equiv \sigma_j,
\qquad j=0,1,2,3.
$$
These identities play a role in quantum theory \cite{1}.
An extension is
$$
\exp\left(-\frac12 i\pi
(\sigma_{j_1} \otimes \sigma_{j_2} \otimes \cdots \otimes \sigma_{j_n} 
- I_{2^n})\right) \equiv 
\sigma_{j_1} \otimes \sigma_{j_2} \otimes \cdots \otimes \sigma_{j_n},
\qquad j_k=0,1,2,3
$$ 
where the $2^n \times 2^n$ matrices 
$\sigma_{j_1} \otimes \sigma_{j_2} \otimes \cdots \otimes \sigma_{j_n}$
are elements of the Pauli group.
Thus we ask the question: find all $n \times n$ matrices
$A$ over $\mathbb C$ and $z \in {\mathbb C}$ such that
$$
e^{z(A-I_n)} = A.
\eqno(1)
$$
Equation (1) can also be written in the form $e^{zA}=e^z A$.
The trivial solution is given by $A=I_n$ with $z$ arbitrary.
Since $\det(e^M)\equiv e^{{\rm\; tr}(M)}$ for any $n \times n$ matrix
$M$ over $\mathbb C$ we obtain
$$
\det(e^{z(A-I_n)}) = e^{{\rm\; tr}(z(A-I_n))} =
e^{z({\rm\; tr}(A)-n)} = \det(A).
$$
Since $\exp(z(\mbox{tr}(A)-n))$ is nonzero we can conclude
that $A$ must be invertible. Note that if $A$ and $B$ are
similar matrices, then $B$ satisfies (1) when $A$ satisfies (1) and
vice versa.
An important special case for a solution of equation (1) can be 
given at once. Let $B$ be an $n \times n$ matrix with $B^2=I_n$.
Then 
$$
\exp\left(\frac12 (2k+1)i\pi(B-I_n)\right) \equiv B\qquad\forall k\in\mathbb{Z}\,.
\eqno(2)
$$
The proof is based on the identity $(z \in {\mathbb C})$
$$
e^{zB} \equiv \cosh(z)I_n + \sinh(z)B
$$
for any $n \times n$ matrix $B$ with $B^2=I_n$. Setting $z=i\pi/2+ik\pi$ we 
obtain the identity (2) utilizing that $\sin(\pi/2+k\pi)=(-1)^k$ and $\cos(\pi/2+k\pi)=0$.
No other solution exists when $B^2=I_n$.
\newline

We investigate first normal matrices and then nonnormal matrices.
Finally a number of applications are provided.

\section{Solutions for Normal Matrices}

We solve the problem under the assumption that $A$ is a normal matrix,
i.e. $AA^*=A^*A$. Let $U$ be the unitary matrix that diagonalizes $A$,
i.e. 
$$
U^{-1}AU = D \equiv \mbox{diag}(\lambda_1,\dots,\lambda_n)
$$
where $\lambda_j$ are the eigenvalues of $A$. Then from
$e^{z(A-I_n)}=A$ we find
$$
U^{-1}e^{z(A-I_n)}U = U^{-1}AU \quad \Rightarrow \quad
e^{zU^{-1}(A-I_n)U} = U^{-1}AU\,.
$$
Consequently $e^{z(D-I_n)}=D$ and 
$$
e^{z(\lambda_j-1)} = \lambda_j, \qquad j=1,\dots,n\,
$$
or equivalently
$$
e^{-z} = \lambda_j e^{-z\lambda_j}, \qquad j=1,\dots,n\,.
$$
Clearly $z=0$ gives $A=I_n$. However, $A=I_n$ does not
constrain $z$. In the following we restrict our discussion
to the case $z\neq 0$ and $A\neq I_n$.\\

The solution of $e^{-z}=\lambda e^{-z\lambda}$ can be given as
$$
\lambda = -\frac1{z}W(-ze^{-z})
$$
where $W$ is (any branch of) the Lambert $W$-function \cite{2,3,4}.
The Lambert $W$-function is defined by $z=W(z)\exp(W(z))$
with the properties 
$$
W(0)=0, \quad W(e)=1, \quad W(-\pi/2)=i\pi/2\,.
$$
Thus for any $z\in\mathbb{C}$ we can construct a normal
matrix $A$ satisfying (1) using the Lambert $W$-function.
In particular, consider $z=-i\omega t$, where $\omega$ is the
frequency and $t$ the time. Then
$$
\lambda = \frac1{i\omega t} W(i\omega t e^{i\omega t})\,.
$$
With $\omega=\pi/2$ we have $e^{i\omega t}=i$ and hence
$\lambda=1$ since the Lambert $W$-function satisfies $W(-\pi/2)=i\pi/2$.
The Lambert $W$-function for matrices has been studied by
Higham \cite{5} and Corliss et al \cite{6}.\\

Now we consider the relationship between eigenvalues of $A$.
Since $e^{-z}=\lambda_je^{-z\lambda_j}$ for all $j\in\{1,\ldots,n\}$.
If $\lambda_j=1$ then the $j$-th equation is satisfied identically.
If $\lambda_j\neq1$ then there must exist $k_j\in\mathbb{Z}$ such that
$$
z=\frac{\ln \lambda_j+2\pi k_j i}{\lambda_j-1}.
$$
Suppose there exists $\lambda_p\neq1$ with $\lambda_p\neq\lambda_j$, then we
also have
$$
z=\frac{\ln \lambda_p+2\pi k_p i}{\lambda_p-1}
$$
for some $k_p\in\mathbb{Z}$. It follows that
$$
k_p=\frac{1}{2\pi i}\left[\frac{\lambda_p-1}{\lambda_j-1}(\ln\lambda_j+2\pi k_j i)-\ln\lambda_p\right]\in\mathbb{Z}.
$$
We have the following cases:
\begin{enumerate}
 \item $A=I_n$, $z\in\mathbb{C}$
 \item $A$ has $r\neq 0$ distinct eigenvalues $\lambda_1\neq 1$, $\lambda_2\neq 1$, \ldots, $\lambda_r\neq 1$;
$$
z=(\ln \lambda_1+2\pi k i)/(\lambda_1-1)
$$
for some $k\in\mathbb{Z}$ and for all $j\in\{1,\ldots,r\}$
$$
\frac1{2\pi i}\left[\frac{\lambda_j-1}{\lambda_1-1}(\ln\lambda_1+2\pi k i)-\ln\lambda_j\right]\in\mathbb{Z}
$$
and any remaining eigenvalues are 1.
\end{enumerate}

Thus $A$ and $z$ satisfy (1) if and only if one of the above cases hold.

\section{Solutions for Nonnormal Matrices}

Note that also some nonnormal matrices $N$ can satisfy the
condition that $N^2=I_n$, so that the solution (2) holds.
Consider for example
$$
N = \pmatrix { 1 & \epsilon \cr 0 & -1 } \equiv 
\sigma_3 + \pmatrix { 0 & \epsilon \cr 0 & 0 }
$$
with $\epsilon \ne 0$. Thus $N^*N \ne NN^*$.
Another example is the matrix
$$
M = \pmatrix { 1 & 0 & \epsilon \cr 0 & -1 & 0 \cr 0 & 0 & -1 } 
$$
with $\epsilon \neq 0$.
\newline

In general, if $A$ is a fixed $n\times n$ matrix, 
let $\{I_n,A,A^2,\ldots,A^r\}$ be the largest
linearly independent set constructed from powers
of $A$ (by the Cayley-Hamilton theorem $r\leq n-1$)
then there exists $c_j(z)$ ($j=0,1,\ldots,r$) such that
$$
\sum_{j=0}^{r}c_j(z)A^j = e^{zA} = e^{z}A.
$$
Thus any solutions $z$ satisfy
$$
c_j(z)=0,\quad j=0,2,3,\ldots,n\qquad c_1(z)=e^z.
$$
The case $N^2=I_n$ above follows as a special case when $r=1$.
\newline

Similar to the normal matrices case in the previous section,
the eigenvalues of $A$ obey the relations below. However,
in these cases additional constraints on $U$ are necessary.
\newline

Assume $A$ satisfies (1), then we have the following cases:
\begin{enumerate}
 \item $A=I_n+U$, $z\in\mathbb{C}$; for some strictly upper triangular matrix $U\neq 0$
 \item $A$ has $r\neq 0$ distinct eigenvalues $\lambda_1\neq 1$, $\lambda_2\neq 1$, \ldots, $\lambda_r\neq 1$;
$$
z=(\ln \lambda_1+2\pi k i)/(\lambda_1-1)
$$
for some $k\in\mathbb{Z}$ and for all $j\in\{1,\ldots,r\}$
$$
\frac{1}{2\pi i}\left[\frac{\lambda_j-1}{\lambda_1-1}(\ln\lambda_1+2\pi k i)-\ln\lambda_j\right]\in\mathbb{Z}
$$
and any remaining eigenvalues are 1.
\end{enumerate}

Thus if $A$ and $z$ satisfy (1) then one of the above cases hold.

\section{Applications}

Let $\hat H$ be the Hamilton operator acting in a finite
dimensional Hilbert space ${\mathbb C}^n$. Thus $\hat H$ would
be an $n \times n$ hermitian matrix. The solution of the Schr\"odinger equation 
is given by
$$
|\psi(t)\rangle = \exp(-i\hat Ht/\hbar)|\psi(0)\rangle\,.
$$
Thus if $\hat H=\hbar\omega K$ with $K^2=I_n$ we have
$$
\exp(-i\omega t K) = \cos(\omega t)I_n - i\sin(\omega t)K
$$
where we utilized that $\cosh(-i\omega t) \equiv \cos(\omega t)$,
$\sinh(-i\omega t)\equiv -i\sin(\omega t)$. 
Thus to satisfy the equation $e^{zA}=e^zA$ we have to set
$\omega t=\pi/2$. Then we obtain
$$
\exp(-i\frac{\pi}{2} K) = e^{-i\pi/2}K
$$
with $e^{-i\pi/2}=-i$. 
\newline

As an example consider the triple spin Hamilton operator 
(Steeb \cite{7})
$$
\hat H = \hbar\omega (\sigma_1 \otimes \sigma_3 \otimes \sigma_2).
$$
Since 
$(\sigma_1 \otimes \sigma_3 \otimes \sigma_2)^2=I_2 \otimes I_2 \otimes I_2$
we obtain 
$$
e^{-i\hat Ht/\hbar} = 
e^{-i\omega t(\sigma_1 \otimes \sigma_3 \otimes \sigma_2)}
= I_8 \cos(\omega t) - i\sin(\omega t)(\sigma_1 
\otimes \sigma_3 \otimes \sigma_2)\,.
$$
If $\omega t=\pi/2$, then $\cos(\pi/2)=0$, $\sin(\pi/2)=1$
and we obtain 
$$
\exp(-i\frac{\pi}2 (\sigma_1 \otimes \sigma_3 \otimes \sigma_2))
= -i\sigma_1 \otimes \sigma_3 \otimes \sigma_2
$$
where $e^{-i\pi/2}=-i$.
\newline

Many quantum gates (\cite{8,9,10,11}) such as the Hadamard gate 
$U_H$ and the CNOT-gate
$$
U_H = \frac1{\sqrt2} \pmatrix { 1 & 1 \cr 1 & -1 }, \qquad
U_{CNOT} = \pmatrix { 1 & 0 & 0 & 0 \cr 0 & 1 & 0 & 0 \cr
                      0 & 0 & 0 & 1 \cr 0 & 0 & 1 & 0 }
$$
and Swap gate
$$
U_{swap} = \pmatrix { 1 & 0 & 0 & 0 \cr 0 & 0 & 1 & 0 \cr
                      0 & 1 & 0 & 0 \cr 0 & 0 & 0 & 1 }
$$
satisfy the condition that the square is the identity matrix.
\newline

If the $n \times n$ matrices $X$ and $Y$ satisfy $X^2=I_n$,
$Y^2=I_n$, then $X \otimes Y$ and $X \oplus Y$ satisfy
$(X \otimes Y)^2=I_{n^2}$ and $(X \oplus Y)^2=I_{2n}$, where
$\oplus$ denotes the direct sum.
An application would be the Pauli spin matrices, for 
example $\sigma_1 \otimes \sigma_2$, $\sigma_1 \oplus \sigma_2$
or $\sigma_1 \otimes \sigma_3 \otimes \sigma_2$,
$\sigma_1 \oplus \sigma_3 \oplus \sigma_2$. The elements
$\sigma_1 \otimes \sigma_2$, 
$\sigma_1 \otimes \sigma_3 \otimes \sigma_2$ are elements of
the Pauli group. The $n$-qubit Pauli group ${\cal P}_n$ is defined by
$$
{\cal P}_n := \{ \, I_2, \sigma_1, \sigma_2, \sigma_3 \, \}^{ \otimes n}
\otimes \{ \, \pm 1, \pm i \, \}\,.
$$
The $n$-qubit Pauli group ${\cal P}_n$ is of order $4^{n+1}$.
\newline

\section{Conclusion}

We solved the matrix equation $\exp(zA)=\exp(z)A$ for normal 
matrices. We have also shown that solutions for nonnormal 
matrices exist.\\

\strut\\
{\bf Acknowledgment}\\

The authors are supported by the National Research Foundation (NRF),
South Africa. This work is based upon research supported by the National
Research Foundation. Any opinion, findings and conclusions or recommendations
expressed in this material are those of the author(s) and therefore the
NRF do not accept any liability in regard thereto.

\strut\hfill

\end{document}